\documentclass[11pt]{article}
\usepackage{graphicx}
\usepackage{latexsym}
\begin{document}
%\pagenumbering{roman}
%\pagenumbering{Roman}
%\pagenumbering{alph}
%\pagenumbering{Alph}
%\setcounter{page}{1}
\setcounter{equation}{0}
\setcounter{figure}{0}
\def\mib#1{\mbox{\boldmath $#1$}}

%%%%%%%%%%%%%%%%%%%%%%%%%%%%%%%%%%%%%%%%%%%%%%%%%%%%%%%%%%%%
%\begin{titlepage}
\begin{flushright}
%31 May, 2004
\end{flushright}

\vskip1cm
\begin{center}
{\Large{\bf Full One-Loop Radiative Corrections to the Asymmetry Parameter
of Polarised Neutron Beta Decay} }
\end{center}

\vskip0.5cm
\begin{center}
{\large {\bf Masataka Fukugita}}
\end{center}

\vskip0.2cm
\begin{center}
{\it Institute for Cosmic Ray Research, University of Tokyo}

{\it Kashiwa  277-8582, Japan}

\begin{center}
and
\end{center}

\begin{center}
{\large {\bf Takahiro Kubota}}
\end{center}

\vskip0.2cm
{\it Graduate School of Science, Osaka University,

Toyonaka, Osaka 560-0043, Japan}
\end{center}

\vskip1cm
\begin{abstract}
%\begin{quote}
We present a calculation of full one-loop radiative corrections,
including the constant term,
to the asymmetry parameter of polarised neutron beta decay.
This gives the radiative correction to the axial coupling constant
$g_A$ extracted from the beta asymmetry so that it ties to $g_A$ that
appears in neutron decay lifetime in a consistent renormalisation
scheme. We find that the ratio of axial-vector to vector
couplings determined from the beta asymmetry,
after taking account of the outer radiative correction, is
related to the bare value as $G_A/G_V=1.0012 G_A^0/G_V^0$.
\end{abstract}

\vskip0.5cm


\vfill\eject

Beta decay asymmetry of the polarised neutron has been used to determine
the axial vector coupling constant $g_A$ of the nucleon. One-loop
radiative corrections to the asymmetry parameter has been calculated
by several authors \cite{shann}-\cite{glueck},
%Garc\'\i a \& Maya and Gl\"uck and T\'oth \cite{garcia,glueck},
but their calculations do not include
the constant term, or so-called inner corrections, which require a
special care in the treatment of the UV divergence of the
radiative correction. This makes the identification of $g_A$ extracted from
$\beta$ decay asymmetry with that which appears in the nucleon
beta decay rate ambiguous\footnote{It is included in Sirlin's
proof \cite{sirlin} that the inner radiative corrections can
be factored out for the processes concerning the unpolarised 
neutron. As far as the authors know there is no proof which shows
that the inner radiative correction for the polarised neutron
and that for the unpolarised neutron should agree.}. The
radiative correction to beta decay is UV divergent and
it is rendered finite only with the use of Weinberg-Salam's theory
of electroweak interaction. The complication arises from the fact that
one must deal with quarks in electroweak theory, and one
must continue to calculations with hadrons at low energies\cite{sirlin2}.

In this paper we calculate one-loop radiative corrections including
the inner correction:
hereby the coupling constant $g_A$ that appears in the $\beta$ asymmetry
parameter is unambiguously tied to that in the decay rate
in a consistent renormalisation scheme.
The key point of the calculation is the clarification of the
universal and non-universal UV divergent parts by using the current
algebra technique and the proof that the same combinations of the
renormalisation factors appear in the beta asymmetry as in the beta decay
rate. The separation of the UV divergences into universal and
non-universal parts  was done first by Abers et al. \cite{abers}
for the Fermi transition of nuclear beta decay, and then used by
Sirlin \cite{sirlin,sirlin2} to develop
a practical scheme of one-loop radiative corrections for the
$0^+\rightarrow 0^+$ transition.
The scheme was extended to the Gamow-Teller transition
by our recent publication (Paper I) \cite{papI}.
The present work is an application of the formalism developed
in Paper I. We are content with the outline of the calculations
in this paper, since the bulk of lengthy calculations are parallel to the
ones
presented in Paper I. We refer the readers who are interested in
technical details of calculations to Paper I.

The tree amplitude for beta decay of the polarised neutron is given by
\begin{eqnarray}
{\cal M}^{(0)}= \frac{G_{V}}{\sqrt{2}}
\left [\bar u_e(\ell)\gamma ^{\lambda }(1-\gamma ^{5})
v_{\nu}(p_\nu)\right ]
\left [\bar u_{p}(p_{2})W_{\lambda }
(p_{2}, p_{1})
\frac{1}{2}\left(1+s\gamma^5\gamma^\mu n_\mu\right)
u_{n}(p_{1})\right ]\ ,
\label{eq:born}
\end{eqnarray}
where $G_{V}=G_{F}{\rm cos}\theta _{C}$ with
the universal Fermi coupling $G_{F}$ and the Gell-Mann-L\'evy-Cabibbo
angle $\theta _{C}$,
$n_\mu$ is the polarisation vector of the neutron with
$n^{2}=-1$, $n\cdot p_{1}=0$, and
$s=\pm 1$, and
$W_\lambda(p_2,p_1)$ is a general form of the weak vertex of hadrons
and reads
\begin{equation}
W_\lambda(p_2,p_1)=\gamma_\lambda(f_V-g_A\gamma^5)
\end{equation}
at low energies.
We retain $f_V=1$ to trace the vector coupling in the calculation.
The spinors of the neutron, proton, electron and
antineutrino are denoted by
$u_{n}$, $u_{p}$, $u_{e}$, and $v_{\nu}$, respectively,
with the momenta specified in parentheses.
After spin summation and integration over $\vec{p}_\nu$
the amplitude square reads
\begin{eqnarray}
\sum _{{\rm spin}}\vert {\cal M}^{(0)} \vert ^{2}
=16 G_{V}^{2} m_{n} m_{p} E_{\nu}\left[
(f_{V}^{2}+3g_{A}^{2})E+2(f_{V}g_{A}-g_{A}^{2})
(\vec{n}\cdot \vec{\ell})
\right],
\end{eqnarray}
where $E$ and $E_\nu$ are energies of the electron and the antineutrino.
Therefore, the asymmetry parameter is given by
\begin{equation}
A=\frac{2(f_Vg_{A}-g_{A}^{2})}{f_V^{2}+3g_{A}^{2}},
\end{equation}
the electron velocity factor $\beta=\vert \vec{\ell}\vert /E$
being removed as a convention.

To evaluate full one loop corrections, we divide the integration region of
the virtual gauge bosons into long- and short-distance parts
\cite{sirlin2}:
\begin{eqnarray}
({\rm i}) \hskip0.5cm 0 < \vert k \vert ^{2} < M^{2},
\hskip1cm
({\rm ii}) \hskip0.5cm M^{2} <\vert k \vert ^{2} < \infty ,
\label{eq:region}
\end{eqnarray}
where $k$ is the momentum of the virtual
gauge bosons, and the mass scale $M$, introduced by hand, divides
the low- and high-energy regimes and is supposed
to lie between the proton-neutron masses ($m_{p}$ and $m_{n}$) and
the $W$ and $Z$ boson masses, $m_{W}$ and $m_{Z}$.
Old-fashioned four-Fermi interactions are applied to the proton and
neutron in region (i), and
the mass scale $M$ is regarded as the ultraviolet cutoff
of the QED (i.e., purely photonic) correction. In region (ii),
electroweak theory is used for quarks and leptons, and
$M$ is the mass scale that describes the onset of the asymptotic behaviour.
The concern is to connect the results in (i) and (ii) smoothly.
Abers et al.  \cite{abers} proved that the logarithmic divergences
that are proportional to $f_V^2$
are universal for the Fermi transition on the basis of
the conserved vector current
with the use of the current algebra technique.
The same was proven for the $g_A^2$ terms for 
the Gamow-Teller transition for which 
current conservation is broken only with soft operators \cite{sirlin3,
papI}. This guarantees smooth connection of the logarithmic divergence
for the corrections of $f_V^2$ and $g_A^2$. There appear, however,
intereference terms of the order $f_Vg_A$, for which logarithmic divergences
depend on the model of hadrons. Marciano \& Sirlin \cite{marciano}
proposed the prescription to
evaluate the high and low energy contributions separately by
rendering the UV divergence in the low-energy contribution milder
by taking account of form factors of hadrons. We follow the same
prescription \cite{marciano,towner}
in the present calculation of the asymmetry parameter.

The diagrams of QED one-loop corrections
are depicted in Figure 1, where
(v) is the vertex correction, (s) is the self energy correction
and (b) is bremsstrahlung.  We write the one-loop amplitude
\begin{equation}
{\cal M'}={\cal M}^{(v)}+{\cal M}^{(s)},
\end{equation}
The bremsstrahlung contribution is added separately.
We consider the static limit for
nucleons, $q^2=(p_1-p_2)^2\ll m_p^2$.
Our calculation is done in the Feynman gauge.

We start with the vertex correction, which is given by
\begin{eqnarray}
{\cal M}^{(v)}&=&
\frac{i}{2\sqrt{2}}G_{V}e^{2}\int \frac{d^{4}k}{(2\pi )^{4}}
\frac{1}{(\ell -k)^{2}-m_{e}^{2}}
\frac{1}{(p_{2}+k)^{2}-m_{p}^{2}}
\frac{1}{k^{2}-\lambda ^{2}}
\nonumber \\
& & \times \bar u_{e}(\ell)\gamma ^{\mu}
\left \{\gamma \cdot (\ell -k)+m_{e}\right \} 
\gamma ^{\lambda }(1-\gamma ^{5})
v_{\nu}(p_\nu)
\nonumber \\
& &\times \bar u_{p}(p_{2})\gamma _{\mu}
\left \{\gamma \cdot (p_{2}+k)+m_{p} \right \}W_{\lambda }(p_{2}+k, p_{1})
(1+s \gamma ^{5}\gamma \cdot n) u_{n}(p_{1}),
\label{eq:virtual}
\end{eqnarray}
where $\lambda $ is the photon mass to regulate
the infrared divergence.
Using identities,
\begin{eqnarray}
\bar u_e(\ell)\gamma^\mu\left \{\gamma \cdot (\ell-k )+m_{e}\right \}
&=&
\bar u_e(\ell)\left \{(2\ell-k )^{\mu}+i\sigma ^{\mu \nu}k_{\nu}\right \},
\label{eq:bunkai1}
\\
\bar u_{p}(p_{2})\gamma _{\mu}\left \{ \gamma \cdot (p_{2}+k)+m_{p}\right \}
&=&
\bar u_{p}(p_{2})\left \{(2p_{2}+k)_{\mu}-i\sigma _{\mu \nu}k^{\nu}
\right \},
\label{eq:bunkai2}
\end{eqnarray}
we decompose (\ref{eq:virtual}) into three parts,
\begin{eqnarray}
{\cal M}^{(v)}={\cal M}^{(v1)}+{\cal M}^{(v2)}+
{\cal M}^{(v3)}.
\end{eqnarray}
Here ${\cal M}^{(v1)}$ picks up the product of
$(2\ell-k )^{\mu}$ in (\ref{eq:bunkai1}) and
$(2p_{2}+k)_{\mu}$ in (\ref{eq:bunkai2}), and at the same time
$W_{\lambda }(p_{2}+k, p_{1})$ is replaced by
$W_{\lambda }(p_{2}, p_{1})$.
It  has the same gamma matrix structure
as the Born term (\ref{eq:born}), and is then written
as a multiplicative correction factor.  This correction
has both UV and IR divergences and depends on the electron
velocity.
The UV divergence in  ${\cal M}^{(v1)}$ is cancelled by 
that in the self energy correction of ${\cal M}^{(s)}$,
\begin{eqnarray}
{\cal M}^{(s)}&=&\left \{
\sqrt{Z_{2}(m_{e})}- 1 + \sqrt{Z_{2}(m_{p})}-1 \right \}
{\cal M}^{(0)},
\label{eq:mc}
\end{eqnarray}
and the IR divergence,
along with that arising from ${\cal M}^{(s)}$, is
cancelled when the contribution of bremsstrahlung
${\cal M}^{(b)}$ is added.

The term ${\cal M}^{(v2)}$ represents the combination of
$i\sigma ^{\mu \nu}k_{\nu}$ in (\ref{eq:bunkai1}) and
$(2p_{2}+k)_{\mu}$ in (\ref{eq:bunkai2}),  and
$W_{\lambda }(p_{2}+k, p_{1})$ is again replaced with
$W_{\lambda }(p_{2}, p_{1})$. This term is UV and IR finite,
but gives an electron-velocity dependent factor.
In the static limit of the nucleon the correction is
also a multiplication on the tree amplitude.

A straightforward calculation yields
\begin{eqnarray}
& &\sum _{{\rm spin}}
\left \{
({\cal M}^{(v1)}+{\cal M}^{(v2)}){\cal M}^{(0)*} + {\rm c.c.}
\right \} +\int\frac{d^3\vec{k}}{(2\pi)^3 2\omega}
\sum_{\rm spin}\left|{\cal M}^{(b)}\right|^2
\frac{(E_{0}-E-\omega)}{(E_{0}-E)}
\nonumber \\
&=& 16 G_{V}^{2} m_{n}m_{p} E_{\nu}
\frac{e^{2}}{8\pi ^{2}}\Bigg [\left ( g(E)-\frac{3}{4}\right )
\left (f_V^{2}+3g_{A}^{2}\right )E
\nonumber \\
& & \hskip3cm +2\left ( \hat g(E)-\frac{3}{4}\right )\left (f_V
g_{A}-g_{A}^{2}\right )
s(\vec{n}\cdot\vec{\ell})
\Bigg ],
\label{eq:311}
\end{eqnarray}
where $g(E, E_{0})$ is the conventional $g$ function that
appears in the radiative
correction for the beta decay rate and is defined with an additional constant
3/4 \cite{kinoshitasirlin,sirlin}.
\begin{eqnarray}
g(E, E_0)
&=&
3\  {\rm ln} \left( {m_p \over m_e}\right)
- {3 \over 4} + {4 \over \beta } L
\left({{2\beta }\over {1+\beta }}\right) \nonumber \\
& &  + 4 \left (\frac{1}{\beta }{\rm tanh}^{-1} \beta  -1 \right )
\left[{{E_0 -E} \over {3E}} - {3 \over 2} + {\rm ln}
{{2(E_0 -E) } \over m_e}\right]\nonumber \\
& & + { 1 \over \beta } {\rm tanh}^{-1} \beta
\left \{2(1+\beta ^2)
+ {{(E_0 -E)^2} \over {6E^2}} - 4 {\rm tanh}^{-1} \beta  \right \}~,
\label{eq:3.6.18}
\end{eqnarray}
where $E_0$ is the end point energy of the electron, and
$\hat g(E, E_{0})$ is a similar function for the spin-dependent term,
\begin{eqnarray}
\hat g(E, E_0)
&=&
3\  {\rm ln} \left( {m_p \over m_e}\right)
- {3 \over 4} + {4 \over \beta } L
\left({{2\beta }\over {1+\beta }}\right) \nonumber \\
& &  + 4 \left (\frac{1}{\beta }{\rm tanh}^{-1} \beta  -1 \right )
\left[{{E_0 -E} \over {3E\beta^2}} - {3 \over 2} +
{{(E_0 -E)^2} \over {24E^2\beta^2}}
+ {\rm ln} {{2(E_0 -E) } \over m_e}\right]\nonumber \\
& & + { 4 \over \beta } {\rm tanh}^{-1} \beta
\left (1- {\rm tanh}^{-1} \beta \right)~.
%\label{eq:3.6.18}
\end{eqnarray}
Here
\begin{eqnarray}
L(z)=\int _{0}^{z}\frac{dt}{t}{\rm log}(1-t),
%=-\sum _{k=1}^{\infty} \frac{z^{k}}{k^{2}}
\end{eqnarray}
is the Spence function. We also define $\hat g(E, E_{0})$
with an additional constant 3/4 as a convention.
These are the outer corrections, which we may write
\begin{equation}
\delta_{\rm out}=\frac{e^2}{8\pi^2} g(E, E_{0}),~~~~~~
\hat\delta_{\rm out}=\frac{e^2}{8\pi^2} \hat g(E, E_{0})
\end{equation}
and agree
with the formulae given by Shann \cite {shann} and by
Garc\'\i a and Maya \cite{garcia}. They also agree with
the expression derived by Yokoo et al. \cite{yokoo} up to a constant.

We now consider the remaining term ${\cal M}^{(v3)}$,
\begin{eqnarray}
{\cal M}^{(v3)}&=&
\frac{i}{2\sqrt{2}}G_{V}e^{2}\int \frac{d^{4}k}{(2\pi )^{4}}
\frac{1}{(\ell-k )^{2}-m_{e}^{2}}
\frac{1}{(p_{2}+k)^{2}-m_{p}^{2}}
\frac{1}{k^{2}-\lambda ^{2}}
\nonumber \\
& & \times \bar u_{e}(\ell)
\left \{(2\ell-k )^{\mu}+i\sigma ^{\mu \nu}k_{\nu}
\right \} \gamma ^{\lambda }(1-\gamma ^{5})
v_{\nu}(p_\nu)
\nonumber \\
& & \times \bar u_{p}(p_{2})
R_{\mu \lambda }(p_{2}, p_{1},k)(1+s\gamma ^{5}\gamma_ \cdot n)
u_{n}(p_{1}),
\label{eq:ma3}
\end{eqnarray}
where
\begin{eqnarray}
R_{\mu \lambda}(p_{1}, p_{2}, k)&= &
(2p_{2}+k)_{\mu}
\left \{ W_{\lambda }(p_{2}+k, p_{1})-W_{\lambda }(p_{2}, p_{1})
\right \}
% \nonumber \\
% & & \\
-i \sigma _{\mu \nu} k^{\nu}
W_{\lambda }(p_{2}+k, p_{1})
\label{eq:rmulambda}
\nonumber \\
%& &
%\\
&\simeq& -i\sigma _{\mu \nu} k^{\nu} W_{\lambda }(p_{2}+k, p_{1})
\nonumber \\
& \simeq &
-i \sigma _{\mu \nu} k^{\nu}\gamma_\lambda(f_V-g_A\gamma^5)\ ,
\end{eqnarray}
in the approximation of the point nucleons.
It is only this ${\cal M}^{(v3)}$ term that
depends on the details of the hadronic part of the
weak current. It is clear from the powers of $k$
that this term is IR convergent, whereas it is UV divergent.

A straightforward  calculation leads to
\begin{eqnarray}
& & \sum _{{\rm spin}}\left \{
{\cal M}^{(v3)}{\cal M}^{(0)*}+
{\cal M}^{(v3)*}{\cal M}^{(0)}
\right \}
\nonumber \\
& & =
16G_{V}^{2}m_{n}m_{p}
\Bigg[
\left\{\frac{3}{2}(f_V^2+f_Vg_A)\log\left(\frac{M}{m_p}\right)^2
       +\left(\frac{3}{4}f_V^2+\frac{9}{4}f_Vg_A\right)\right\}
            (EE_{\nu}+\vec{\ell}\cdot\vec{p_\nu}) \cr
& & + \left\{\frac{3}{2}(g_A^2+f_Vg_A)\log\left(\frac{M}{m_p}\right)^2
       +\left(\frac{7}{4}g_A^2+\frac{5}{4}f_Vg_A\right)\right\}
            (3EE_{\nu}-\vec{\ell}\cdot\vec{p_\nu})  \cr
& & + 2s\left\{ \left (\frac{3}{4}f_V^2+\frac{3}{4}g_A^2+\frac{3}{2}f_Vg_A
                       \right )
                      \log\left(\frac{M}{m_p}\right)^2
                      +\left(\frac{5}{8}f_V^2+\frac{9}{8}g_A^2+\frac{5}{4}
                      f_Vg_A\right)\right\}
\nonumber \\
& &   \hskip3cm \times
 \left ( E(\vec{n}\cdot\vec{p_\nu})+E_\nu (\vec{n}\cdot\vec{\ell})
            \right )   \cr
& & + 2s\left\{ \frac{3}{2}(g_A^2+f_Vg_A)\log\left(\frac{M}{m_p}\right)^2
    +\left(\frac{7}{4}g_A^2+\frac{5}{4}f_Vg_A\right)\right\}
            \left ( E(\vec{n}\cdot\vec{p_\nu})-E_\nu (\vec{n}\cdot
\vec{\ell}
            )\right )
\Bigg]\ .
\end{eqnarray}

We use the current algebra technique to classify logarithmic divergences
into those that have universal coefficients irrespective of the model
of hadrons and those that are model dependent. 
Repeating the same calculation as in Paper I but including the spin
projection operator, we find that
$(3/2) f_V^2 \log(M/m_p)^2$ in the first,
$(3/2) g_A^2 \log(M/m_p)^2$ in the second,
$(3/2)f_Vg_A \log(M/m_p)^2$ in the third and $(3/2)g_A^2 \log(M/m_p)^2$ in
the
fourth curly brackets are universal. This observation tells us that the
terms are summarised as,
\begin{eqnarray}
& &  \sum _{{\rm spin}}\left \{
{\cal M}^{(v3)}{\cal M}^{(0)*}+
{\cal M}^{(v3)*}{\cal M}^{(0)}
\right \}
\nonumber \\
& & =16G_{V}^{2}m_{n}m_{p} \Bigg[
f_V^2\left( {\delta_{\rm in}^{\rm F}}'+\frac{3}{4}\cdot
\frac{e^2}{8\pi^2}\right)
(EE_{\nu}+\vec{\ell}\cdot\vec{p_\nu})
%\nonumber \\
%& &
+g_A^2\left( {\delta_{\rm in}^{\rm
GT}}'+\frac{3}{4}\cdot \frac{e^2}{8\pi^2}\right)
(3EE_{\nu}-\vec{\ell}\cdot\vec{p_\nu})\cr
& &+2sf_Vg_A\left(\frac{1}{2}{\delta_{\rm in}^{\rm F}}'
  +\frac{1}{2}{\delta_{\rm in}^{\rm
GT}}'+\frac{3}{4}\cdot \frac{e^2}{8\pi^2}\right)
  \left \{ E(\vec{n}\cdot\vec{p_\nu})+E_\nu (\vec{n}\cdot\vec{\ell})\right
\}
    \cr
& &+2s g_A^2\left( {\delta_{\rm in}^{\rm GT}}'+
              \frac{3}{4}\cdot \frac{e^2}{8\pi^2}\right)
\left \{ E(\vec{n}\cdot\vec{p_\nu})-E_\nu (\vec{n}\cdot\vec{\ell})\right \}
\Bigg]\ .
\label{eq:a3}
\end{eqnarray}
Adding the tree term and after integration over $p_\nu$
\begin{eqnarray}
& & \hskip-1cm \sum _{{\rm spin}} \left|{\cal M}^{(0)}\right|^2+
\sum _{\rm spin} \left \{
{\cal M}^{(v3)}{\cal M}^{(0)*}+
{\cal M}^{(v3)*}{\cal M}^{(0)}
\right \}
\nonumber \\
& & =16G_{V}^{2}m_{n}m_{p} E_{\nu} \left (
1+\frac{3}{4}\cdot \frac{e^{2}}{8\pi ^{2}}
\right )
\Bigg[
\left\{f_V^2\left(1+ {\delta_{\rm in}^{\rm F}}' \right)
+ 3g_A^2\left(1+ {\delta_{\rm in}^{\rm GT}}' \right)\right\}E \cr
& &+2s\left\{f_Vg_A\left(1+ \frac{1}{2}{\delta_{\rm in}^{\rm F}}'
+ \frac{1}{2}{\delta_{\rm in}^{\rm GT}}' \right)
-g_A^2\left(1+ {\delta_{\rm in}^{\rm GT}}' \right)\right\}
(\vec{n}\cdot\vec{\ell})\Bigg]\ ,
\label{eq:v3summary}
\end{eqnarray}
where
\begin{eqnarray}
{\delta_{\rm in}^{\rm F}}' &=& \frac{e^{2}}{8\pi ^{2}}\left [\frac{3}{2}
\log\left(\frac{M^2}{m_p^2}\right)
+\frac{g_{A}}{f_{V}}\left \{ \frac{3}{2}{\rm log}\left (
\frac{M^{2}}{m_{p}^{2}} \right )+\frac{9}{4}\right \}
\right]\ ,
\label{eq:in-1-F}\\
{\delta_{\rm in}^{\rm GT}}' &=&\frac{e^{2}}{8\pi ^{2}}
\left[\frac{3}{2} \log\left(\frac{M^2}{m_p^2}\right)+1+
\frac{f_{V}}{g_{A}}\left \{
\frac{3}{2}{\rm log}\left (\frac{M^{2}}{m_{p}^{2}}\right )
+\frac{5}{4}\right \}
\right] 
\label{eq:in-1-GT}
\end{eqnarray}
are the inner corrections
for Fermi and Gamow-Teller transitions 
that are defined in Paper I and the factor  $\left [1+(3/4)e^2/8\pi^2
\right ]$
 is to be included in $\delta_{\rm out}$.
In eqs. (\ref{eq:in-1-F}) and (\ref{eq:in-1-GT}) 
the first logarithms are model-independent
and the second with the coefficients $g_A/f_V$ or $f_V/g_A$ are
model dependent.
The correction from ${\cal M}^{(v3)}$ is written
as multiplicative factors on the coupling constants for both Fermi
and Gamow-Teller parts while they are divergent within QED.

The short distance correction from the integration region (ii)
in (\ref{eq:region}) is evaluated using electroweak theory \cite{sirlin2}.
When we consider corrections
relative to muon decay, we only need to consider the box diagrams of
photon (or $Z$) and $W$ exchanges (see Fig. 2 of Paper I).
In order to connect the quark-level amplitudes with hadronic ones,
we assume that the ratio of the tree and
loop amplitudes for beta decays of the assembly of quarks
is the same as that for neutron beta decay
\cite{sirlin2}. This is justified at least for the universal
logarithmic divergent part.
With this prescription the correction amounts to a multiplication
factor
\begin{eqnarray}
\frac{e^{2}}{8\pi ^{2}}\Bigg[\left \{
\frac{3}{2} {\rm log}\left (\frac{m_{W}^{2}}{M^{2}}
\right ) +3\bar Q {\rm log}\left (
\frac{m_{Z}^{2}}{M^{2}}
\right )
+\frac{5}{2{\rm tan}^{4}\theta _{W}}{\rm log}\left (
\frac{m_{Z}^{2}}{m_{W}^{2}}\right )
\right \}
\cr
-\frac{e^{2}}{8\pi ^{2}}\left (
-\frac{3}{2} +\frac{5}{2{\rm tan}^{4}\theta _{W}}\right )
{\rm log}\left (
\frac{m_{Z}^{2}}{m_{W}^{2}}\right )\Bigg]
\vert {\cal M}^{(0)}\vert ^{2}.
\label{eq:wa}
\end{eqnarray}
where $\bar Q=1/6$ is the mean charge of the isodoublet of quarks, and
the second line is the correction that appears in muon decay
and thus subtracted when we consider the radiative correction
relative to muon decay that determines $G_F$.

This electroweak one-loop correction amounts to adding to
${\delta _{{\rm in}}^{\rm F}}'$ extra terms,
\begin{eqnarray}
\delta _{\rm in}^{\rm F}&\equiv &
{\delta _{{\rm in}}^{\rm F}}'+
\frac{e^{2}}{8\pi ^{2}}\left[ \frac{3}{2}
\log\left(\frac{m_{W}^{2}}{M^{2}}\right)
+3{\bar Q} \log\left(\frac{m_{Z}^{2}}{M^{2}}\right)
+\frac{5}{2{\rm tan}^{4}\theta _{W}}{\rm log}\left (
\frac{m_{Z}^{2}}{m_{W}^{2}}\right )\right ]
\nonumber \\
& &-\frac{e^{2}}{8\pi ^{2}}
\left (-\frac{3}{2}+\frac{5}{2{\rm tan}^{4}\theta _{W}}
\right )
{\rm log}\left (
\frac{m_{Z}^{2}}{m_{W}^{2}}\right )
\nonumber \\
&=&
\frac{e^{2}}{8\pi ^{2}}\left[ \frac{3}{2}
\log\left(\frac{m_{Z}^{2}}{m_{p}^{2}}\right)
+3{\bar Q} \log\left(\frac{m_{Z}^{2}}{M^{2}}\right)
+C_{}^{\rm F}\right ]\ ,
\label{eq:summe}
\end{eqnarray}
where the terms proportional to $g_{A}/f_{V}$ are collected in $C_{}^{\rm
F}$,
\begin{eqnarray}
C_{}^{\rm F}= \frac{g_{A}}{f_{V}}
\left \{
\frac{3}{2}{\rm log}\left (
\frac{M^{2}}{m_{p}^{2}}\right )+\frac{9}{4}
\right \}
\end{eqnarray}
and similarly for ${\delta _{{\rm in}}^{\rm GT}}'$,
\begin{eqnarray}
\delta _{\rm in}^{\rm GT}&=&
\frac{e^{2}}{8\pi ^{2}}\left[\frac{3}{2}
\log\left(\frac{m_{Z}^{2}}{m_p^2}\right)
+1+
3{\bar Q}
\log\left(\frac{m_{Z}^2}{M^{2}}\right)
+C_{}^{\rm GT}
\right]\ ,
\label{eq:d-in-gt}
\end{eqnarray}
where
$C_{}^{\rm GT}$ is
\begin{eqnarray}
C_{}^{\rm GT}= \frac{f_{V}}{g_{A}}
\left \{
\frac{3}{2}{\rm log}\left (
\frac{M^{2}}{m_{p}^{2}} \right ) +\frac{5}{4}
\right \}
\label{eq:1258}
\end{eqnarray}
for point nucleons.

We observe that the $M$ dependence (upper cutoff) that appears in
the first term of (\ref{eq:in-1-F}) is cancelled by the
first term in the braces
in (\ref{eq:wa}), which demonstrates a smooth connection from
electroweak theory to effective hadronic theory for the Fermi
transition.
The UV divergence in the term
proportional to $g_A/f_V$, however, fails to cancel against the
divergence with
the coefficient $3\bar Q$, unless $\bar Q$ has a specific
(and unrealistic) value of charge. Marciano and Sirlin
\cite{marciano}
proposed to evaluate the model-dependent long-distance divergence of
the Fermi transition by
rendering it softer introducing nucleon form factors,
and leave the term $3\bar Q\log (m_Z/M)$ as it is taking
$M$ as the mass scale of the onset of the asymptotic behaviour \cite{marciano}.
The same procedure was followed in Paper I for the GT part,
while it was noted that the inclusion of the weak magnetism
is important  to evaluate the long-distance integral,
especially for the GT transition.

The calculation of the long distance contribution is made replacing
\begin{eqnarray}
\gamma^\mu~~\rightarrow~~ \gamma ^{\mu }F_{1}(k^{2})-\frac{i}{2m_{N}}\sigma
^{\mu \nu}
k_{\nu}F_{2}(k^{2}),
\end{eqnarray}
for the electromagnetic vertex, and
\begin{eqnarray}
W_\lambda (p_{2}, p_{1})~~\rightarrow~~ \gamma_\lambda \left \{f_{V}
F_{V}(k^{2})-g_{A} \gamma^5 F_A({k^2})\right \}-\frac{i}{2m_{p}}
\sigma ^{\mu \nu}
k_{\nu}F_{W}(k^{2}),
\end{eqnarray}
in eq. (\ref{eq:virtual}).
 From the form we observed in eq. (\ref{eq:v3summary}) we expect that
the calculation incorporating form factors would give rise to
a result summarised in the same form, while $C^{\rm F}$ and $C^{\rm GT}$ are
modified exactly as in \cite{papI}. Since we have not found
an immediate proof that it should,
we repeated a long calculation as we did in Paper I 
including the spin
projection operator, and confirmed 
the anticipated result. In fact, we obtained
$C^{\rm F}$ and $C^{\rm GT}$ exactly those that appear in the spin
independent part. So we take the result of numerical integral of Paper I,
\begin{eqnarray}
C_{}^{\rm F}=1.751+0.409=2.160,
\\
C_{}^{\rm GT}=0.727+2.554= 3.281,
\label{eq:coeff}
\end{eqnarray}
where the two parts
of numbers represent contributions from
the (V,A) interaction and weak magnetism.
The first number in $C^{\rm F}$ was evaluated by Marciano and Sirlin
\cite{marciano} and by Towner \cite{towner},
and agrees with their results up to slight differences in the input
parameters.

In conclusion the radiative correction to polarised neutron
beta decay to order $O(\alpha)$ is summarised as
\begin{eqnarray}
& & \hskip-1cm \left|{\cal M}^{(0)}\right|^2+\sum _{{\rm spin}}\left \{
{\cal M}'{\cal M}^{(0)*}+
{\cal M}'{\cal M}^{(0)}\right \}
+\int \frac{d^{3}\vec{k}}{(2\pi )^{3}2\omega }\sum _{\rm spin}
\vert {\cal M}^{(b)}\vert ^{2}\frac{(E_{0}-E-\omega )}{(E_{0}-E)}
\nonumber \\
& & =16G_{V}^{2}m_{n}m_{p} E_\nu\Bigg[
\left\{f_V^2\left(1+ {\delta_{\rm in}^{\rm F}}+\delta_{\rm out}\right)
+g_A^2\left(1+ {\delta_{\rm in}^{\rm GT}}+\delta_{\rm
out}\right)\right\}E\cr
& &+2s\left\{f_Vg_A\left(1+\frac{1}{2}{\delta_{\rm in}^{\rm F}}
  +\frac{1}{2}{\delta_{\rm in}^{\rm GT}}+\hat\delta_{\rm out}\right)
  -g_A^2\left(1+ {\delta_{\rm in}^{\rm GT}}+\hat\delta_{\rm out}\right)\right\}
  (\vec{n}\cdot\vec{\ell})\Bigg]
%\label{eq:a3}
\end{eqnarray}
in the static nucleon approximation. Hence the asymmetry parameter is
written as
\begin{equation}
A=2\frac{1+\frac{\displaystyle{\alpha}}{\displaystyle{2\pi}}
\hat g(E, E_{0})}{1+\frac{\displaystyle{\alpha}}{\displaystyle{2\pi}}
g(E, E_{0})}~
\frac{\bar f_V\bar g_A-\bar g_A^2}{\bar f_V^2+3\bar g_A^2}\ ,
\label{eq:asymmetry-corr}
\end{equation}
where
\begin{eqnarray}
\bar f_V^2 &=& f_V^2(1+{\delta_{\rm in}^{\rm F}}), \cr
\bar g_A^2 &=& g_A^2(1+{\delta_{\rm in}^{\rm GT}}).
\end{eqnarray}

The denominator of eq. (\ref{eq:asymmetry-corr}) is the combination that
appears in the neutron decay rate.
The energy dependent prefactor ${\cal C}(E)$,
\begin{equation}
1+{\cal C}(E)=\left[{1+\frac{\displaystyle{\alpha}}{\displaystyle{2\pi}}\hat
g(E, E_{0})}\right]/\left[{1+\frac{\displaystyle{\alpha}}{
\displaystyle{2\pi}} g(E, E_{0})}\right]
\end{equation}
is plotted in Figure 2 as a function of the kinetic energy
$T=E-m_e$. The magnitude of $(\alpha/2\pi)g(E, E_{0})$ and 
$(\alpha/2\pi)\hat g(E, E_{0})$
is about 2\%, but
the two corrections nearly cancel in $\hat g(E, E_{0})-g(E, E_{0})$,
leaving
the net outer correction for the asymmetry 
being quite small, of the order of 0.1\%.
For convenience we give a fit to  
${\cal C}(E)$ for neutron beta decay with
\begin{equation}
{\cal C}(E)=-0.00163+0.00210/E+0.000491E,
\end{equation}
where $E$ is in units of MeV. The fit, also displayed
in Figure 2, overlays nearly top on the true function of ${\cal C}(E)$. 

The cancellation also takes place for the inner correction.
After correcting for ${\cal C}(E)$, the axial-vector to
vector coupling ratio extracted from the tree level formula
is related to its tree-level value as

\begin{eqnarray}
\frac{\bar g_A}{\bar f_V} &=& \left[1+\frac{\alpha}{4\pi}
\left(1+C^{\rm GT}-C^{\rm F}
\right)\right]\left(\frac{g_A}{f_V}\right) \cr
&=& 1.0012~\frac{g_A}{f_V}.
\label{eq:couplingchg}
\end{eqnarray}
The dominant part of the inner correction, including $\log m_Z/m_p$ cancels
in
$\delta_{\rm in}^{\rm F}-\delta_{\rm in}^{\rm GT}$, and the net correction
is of the order of 0.1\% for $g_A/f_V$ (which is usually denoted as
$\lambda\equiv G_A/G_V=-g_A/f_V$).

The Particle Data Group \cite{pdg} gives a value $g_A=1.2670\pm0.0030$.
This is obtained by averaging 5 values reported in the literature,
one of which \cite{abele} is obtained including the outer radiative
correction
(with the inner correction discarded),
and others are results that
do not include radiative corrections. 
The outer radiative correction reduces the value of $|g_A|$ by about
0.0007, but the scatter among the data from different authors is
0.005 (rms), so systematic errors other than the radiative correction 
dominate the uncertainty of $g_A$.
As we have shown that the inner radiative corrections can be included
into $g_A$ in common irrespective of quantities measured for beta decay, 
it is a matter
of definition whether they are included in $g_A$ or not. If
we define the tree-level axial coupling constant it is related with
the value including the radiative correction by eq. (\ref{eq:couplingchg}).

\vskip1cm
\noindent
{\large {\bf Acknowledgement}}

The work is supported in part by Grants in Aid
of the Ministry of Education. % (Nos 14540265 and 13135215).

%\vfill\eject

\vfil\eject
\noindent
{\bf Figure Captions}

\vskip5mm
\noindent
Fig. 1 radiative corrections to neutron beta decay.

\vskip5mm\noindent
Fig. 2  Outer radiative correction ${\cal C}(T+m_e)$ for the asymmetry
parameter as a function of the kinetic energy of electron. 
The solid curve is ${\cal C}(T+m_e)$, and 
the dotted curve, which overlays nearly exactly on the solid curve,
is fit (37). 

\vfill\eject
\begin{figure}[ht]
%WinTpicVersion2.15
\unitlength 0.1in
\begin{picture}(57.32,23.00)(0.70,-23.40)
% ELLIPSE 1 0 1 0
% 4 740 1464 943 1666 524 1384 524 1384
% 
\special{pn 13}%
\special{sh 0.300}%
\special{ar 740 1064 203 202  0.0000000 6.2831853}%
% LINE 1 0 3 0
% 2 731 1671 731 2208
% 
\special{pn 13}%
\special{pa 731 1271}%
\special{pa 731 1808}%
\special{fp}%
% LINE 1 0 3 0
% 2 578 1330 199 685
% 
\special{pn 13}%
\special{pa 578 930}%
\special{pa 199 285}%
\special{fp}%
% POLYLINE 1 0 3 0
% 4 929 649 929 1375 1344 649 1344 649
% 
\special{pn 13}%
\special{pa 929 249}%
\special{pa 929 975}%
\special{pa 1344 249}%
\special{pa 1344 249}%
\special{fp}%
% STR 2 0 3 0
% 3 686 2271 686 2360 10 41
% $n$
\put(6.8600,-19.6000){\makebox(0,0)[lb]{$n$}}%
% STR 2 0 3 0
% 3 208 542 208 631 10 41
% $p$
\put(2.0800,-2.3100){\makebox(0,0)[lb]{$p$}}%
% STR 2 0 3 0
% 3 830 533 830 622 10 41
% $e^{-}$
\put(8.3000,-2.2200){\makebox(0,0)[lb]{$e^{-}$}}%
% STR 2 0 3 0
% 3 1220 521 1220 610 10 41
% $\bar \nu _{e}$
\put(12.2000,-2.1000){\makebox(0,0)[lb]{$\bar \nu _{e}$}}%
% STR 2 0 3 0
% 3 70 981 70 1070 10 41
% $(p_{2})$
\put(0.7000,-6.7000){\makebox(0,0)[lb]{$(p_{2})$}}%
% STR 2 0 3 0
% 3 956 846 956 936 10 41
% $(\ell )$
\put(9.5600,-5.3600){\makebox(0,0)[lb]{$(\ell )$}}%
% STR 2 0 3 0
% 3 758 1912 758 2002 10 41
% $(p_{1})$
\put(7.5800,-16.0200){\makebox(0,0)[lb]{$(p_{1})$}}%
% ELLIPSE 1 0 1 0
% 4 2241 1463 2444 1665 2025 1382 2025 1382
% 
\special{pn 13}%
\special{sh 0.300}%
\special{ar 2241 1063 203 202  0.0000000 6.2831853}%
% LINE 1 0 3 0
% 2 2232 1669 2232 2207
% 
\special{pn 13}%
\special{pa 2232 1269}%
\special{pa 2232 1807}%
\special{fp}%
% LINE 1 0 3 0
% 2 2079 1329 1700 684
% 
\special{pn 13}%
\special{pa 2079 929}%
\special{pa 1700 284}%
\special{fp}%
% POLYLINE 1 0 3 0
% 4 2439 684 2439 1409 2854 684 2854 684
% 
\special{pn 13}%
\special{pa 2439 284}%
\special{pa 2439 1009}%
\special{pa 2854 284}%
\special{pa 2854 284}%
\special{fp}%
% STR 2 0 3 0
% 3 2187 2269 2187 2359 10 41
% $n$
\put(21.8700,-19.5900){\makebox(0,0)[lb]{$n$}}%
% STR 2 0 3 0
% 3 1701 541 1701 631 10 41
% $p$
\put(17.0100,-2.3100){\makebox(0,0)[lb]{$p$}}%
% STR 2 0 3 0
% 3 2358 540 2358 630 10 41
% $e^{-}$
\put(23.5800,-2.3000){\makebox(0,0)[lb]{$e^{-}$}}%
% STR 2 0 3 0
% 3 4260 531 4260 620 10 41
% $\bar \nu _{e}$
\put(42.6000,-2.2000){\makebox(0,0)[lb]{$\bar \nu _{e}$}}%
% ELLIPSE 1 0 1 0
% 4 3710 1472 3913 1673 3494 1391 3494 1391
% 
\special{pn 13}%
\special{sh 0.300}%
\special{ar 3710 1072 203 201  0.0000000 6.2831853}%
% LINE 1 0 3 0
% 2 3701 1678 3701 2215
% 
\special{pn 13}%
\special{pa 3701 1278}%
\special{pa 3701 1815}%
\special{fp}%
% LINE 1 0 3 0
% 2 3548 1337 3169 692
% 
\special{pn 13}%
\special{pa 3548 937}%
\special{pa 3169 292}%
\special{fp}%
% POLYLINE 1 0 3 0
% 4 3909 674 3909 1400 4323 674 4323 674
% 
\special{pn 13}%
\special{pa 3909 274}%
\special{pa 3909 1000}%
\special{pa 4323 274}%
\special{pa 4323 274}%
\special{fp}%
% STR 2 0 3 0
% 3 3656 2278 3656 2367 10 41
% $n$
\put(36.5600,-19.6700){\makebox(0,0)[lb]{$n$}}%
% STR 2 0 3 0
% 3 3160 549 3160 638 10 41
% $p$
\put(31.6000,-2.3800){\makebox(0,0)[lb]{$p$}}%
% STR 2 0 3 0
% 3 3836 540 3836 629 10 41
% $e^{-}$
\put(38.3600,-2.2900){\makebox(0,0)[lb]{$e^{-}$}}%
% STR 2 0 3 0
% 3 5682 554 5682 644 10 41
% $\bar \nu _{e}$
\put(56.8200,-2.4400){\makebox(0,0)[lb]{$\bar \nu _{e}$}}%
% ELLIPSE 1 0 1 0
% 4 5189 1481 5392 1683 4973 1400 4973 1400
% 
\special{pn 13}%
\special{sh 0.300}%
\special{ar 5189 1081 203 202  0.0000000 6.2831853}%
% LINE 1 0 3 0
% 2 5180 1687 5180 2225
% 
\special{pn 13}%
\special{pa 5180 1287}%
\special{pa 5180 1825}%
\special{fp}%
% LINE 1 0 3 0
% 2 5027 1347 4649 702
% 
\special{pn 13}%
\special{pa 5027 947}%
\special{pa 4649 302}%
\special{fp}%
% POLYLINE 1 0 3 0
% 4 5388 702 5388 1427 5802 702 5802 702
% 
\special{pn 13}%
\special{pa 5388 302}%
\special{pa 5388 1027}%
\special{pa 5802 302}%
\special{pa 5802 302}%
\special{fp}%
% STR 2 0 3 0
% 3 5144 2296 5144 2386 10 41
% $n$
\put(51.4400,-19.8600){\makebox(0,0)[lb]{$n$}}%
% STR 2 0 3 0
% 3 4640 567 4640 657 10 41
% $p$
\put(46.4000,-2.5700){\makebox(0,0)[lb]{$p$}}%
% STR 2 0 3 0
% 3 5289 540 5289 630 10 41
% $e^{-}$
\put(52.8900,-2.3000){\makebox(0,0)[lb]{$e^{-}$}}%
% STR 2 0 3 0
% 3 2809 540 2809 630 10 41
% $\bar \nu _{e}$
\put(28.0900,-2.3000){\makebox(0,0)[lb]{$\bar \nu _{e}$}}%
% STR 2 0 3 0
% 3 600 2821 600 2910 10 41
% (v)
\put(6.0000,-25.1000){\makebox(0,0)[lb]{(v)}}%
% STR 2 0 3 0
% 3 3610 2821 3610 2910 10 41
% (s)
\put(36.1000,-25.1000){\makebox(0,0)[lb]{(s)}}%
% STR 2 0 3 0
% 3 5070 2820 5070 2910 10 41
% (b)
\put(50.7000,-25.1000){\makebox(0,0)[lb]{(b)}}%
% ELLIPSE 1 1 3 0
% 4 1998 1015 2225 1241 1267 504 2025 1355
% 
\special{pn 13}%
\special{ar 1998 615 227 226  1.4920124 1.7569130}%
\special{ar 1998 615 227 226  1.9158534 2.1807541}%
\special{ar 1998 615 227 226  2.3396945 2.6045951}%
\special{ar 1998 615 227 226  2.7635355 3.0284362}%
\special{ar 1998 615 227 226  3.1873766 3.4522773}%
\special{ar 1998 615 227 226  3.6112177 3.7535112}%
% ELLIPSE 1 1 3 0
% 4 3954 988 4130 1163 3845 638 3809 1472
% 
\special{pn 13}%
\special{ar 3954 588 176 175  1.8601796 2.2020600}%
\special{ar 3954 588 176 175  2.4071882 2.7490685}%
\special{ar 3954 588 176 175  2.9541967 3.2960770}%
\special{ar 3954 588 176 175  3.5012052 3.8430856}%
\special{ar 3954 588 176 175  4.0482138 4.3900941}%
% STR 2 0 3 0
% 3 5063 594 5063 684 10 41
% $\gamma $
\put(50.6300,-2.8400){\makebox(0,0)[lb]{$\gamma $}}%
% STR 2 0 3 0
% 3 3629 988 3629 1077 10 41
% $\gamma $
\put(36.2900,-6.7700){\makebox(0,0)[lb]{$\gamma $}}%
% STR 2 0 3 0
% 3 1718 1212 1718 1302 10 41
% $\gamma $
\put(17.1800,-9.0200){\makebox(0,0)[lb]{$\gamma $}}%
% ELLIPSE 1 1 3 0
% 4 830 1760 1572 2497 938 945 370 936
% 
\special{pn 13}%
\special{ar 830 1360 742 737  4.2063140 4.2874499}%
\special{ar 830 1360 742 737  4.3361315 4.4172674}%
\special{ar 830 1360 742 737  4.4659489 4.5470848}%
\special{ar 830 1360 742 737  4.5957663 4.6769023}%
\special{ar 830 1360 742 737  4.7255838 4.8067197}%
% STR 2 0 3 0
% 3 650 1124 650 1214 10 41
% $\gamma $
\put(6.5000,-8.1400){\makebox(0,0)[lb]{$\gamma $}}%
% LINE 1 1 3 0
% 2 5379 1096 5189 702
% 
\special{pn 13}%
\special{pa 5379 696}%
\special{pa 5189 302}%
\special{da 0.070}%
% STR 2 0 3 0
% 3 2100 2821 2100 2910 10 41
% (s)
\put(21.0000,-25.1000){\makebox(0,0)[lb]{(s)}}%
\end{picture}%
\caption{}%radiative correction to neutron beta decay.}
\end{figure}

%%%%%%%%%%%%%%%%%%%%%% fig2 %%%%%%%%%%%%%%%%%
\vfill\eject
\begin{figure}[ht]
\includegraphics*[scale=0.8]{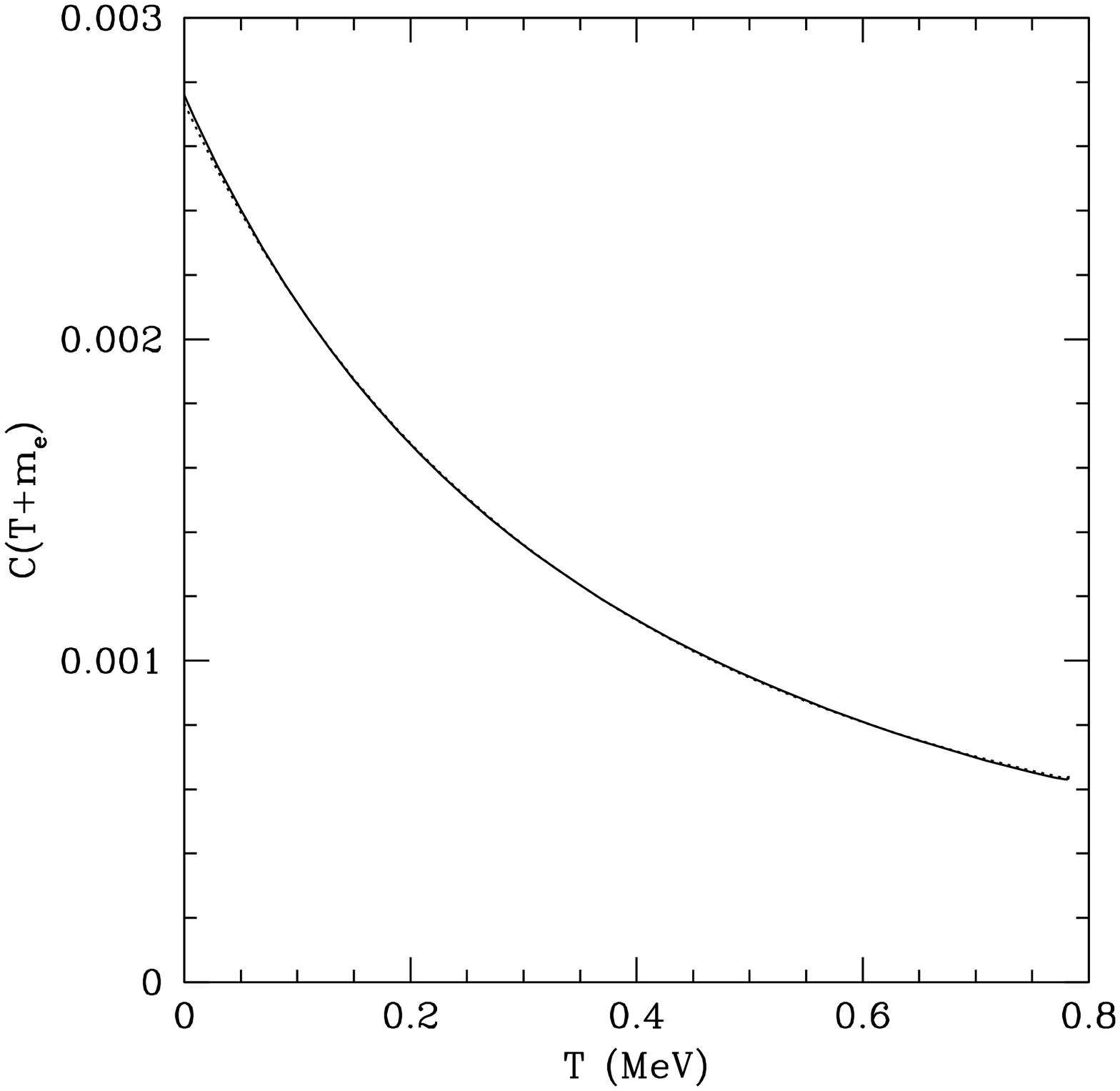}
\caption{}
%Outer radiative correction ${\cal C}(T+m_e)$ 
%for asymmetry as a function of the kinetic 
%energy of the electron. The solid curve is 
%${\cal C}(T+m_e)$, and the dotted curve, which 
%overlays nearly exactly on the solid curve,is fit (37). }
\end{figure}

\end{document}